\documentclass{emulateapj}

\usepackage{url}
\usepackage{epstopdf}
\usepackage{multirow}
\usepackage{amsmath}
\usepackage{natbib}
\usepackage{color}
\usepackage{booktabs}
\usepackage{tabularx}
\usepackage[colorlinks, citecolor=blue, linkcolor=blue, menucolor=blue, urlcolor=blue]{hyperref}

\slugcomment{}

\begin{document}

\title{Ejection of Double knots from the radio core of PKS 1510--089 during the strong $\gamma$-ray flares in 2015}

\author{Jongho Park\altaffilmark{1}, Sang-Sung Lee\altaffilmark{2,3}, Jae-Young Kim\altaffilmark{4}, Jeffrey A. Hodgson\altaffilmark{2}, Sascha Trippe\altaffilmark{1}, Dae-Won Kim\altaffilmark{1}, Juan-Carlos Algaba\altaffilmark{1,2,5}, Motoki Kino\altaffilmark{6,7}, Guang-Yao Zhao\altaffilmark{2}, Jee Won Lee\altaffilmark{2}, and Mark A. Gurwell\altaffilmark{8}}
\affil{$^1$Department of Physics and Astronomy, Seoul National University, Gwanak-gu, Seoul 08826, Republic of Korea; jhpark@astro.snu.ac.kr\\
$^2$Korea Astronomy and Space Science Institute, 776 Daedeok-daero, Yuseong-gu, Daejeon 34055, Republic of Korea\\
$^3$Korea University of Science and Technology, 217 Gajeong-ro, Yuseong-gu, Daejeon 34113, Republic of Korea\\
$^4$Max-Planck-Institut f\"ur Radioastronomie, Auf dem H\"ugel 69, D-53121 Bonn, Germany\\
$^5$Department of Physics, Faculty of Science, University of Malaya, 50603 Kuala Lumpur, Malaysia\\
$^6$National Astronomical Observatory of Japan, 2-21-1 Osawa, Mitaka, Tokyo 181-8588, Japan\\
$^7$Kogakuin University of Technology \& Engineering, Academic Support Center, 2665-1 Nakano, Hachioji, Tokyo 192-0015, Japan\\
$^8$Center for Astrophysics | Harvard \& Smithsonian, 60 Garden Street, Cambridge, MA 02138 USA\\
}
\received{...}
\accepted{...}

\begin{abstract}
\noindent PKS 1510--089 is a bright and active $\gamma$-ray source that showed strong and complex $\gamma$-ray flares in mid-2015 during which the Major Atmospheric Gamma Imaging Cherenkov telescopes detected variable very high energy (VHE; photon energies $>$100 GeV) emission. We present long-term multi-frequency radio, optical, and $\gamma$-ray light curves of PKS 1510--089 from 2013 to 2018, and results of an analysis of the jet kinematics and linear polarization using 43 GHz Very Long Baseline Array data observed between late 2015 and mid-2017. We find that a strong radio flare trails the $\gamma$-ray flares in 2015, showing an optically thick spectrum at the beginning and becoming optically thin over time. Two laterally separated knots of emission are observed to emerge from the radio core nearly simultaneously during the $\gamma$-ray flares. We detect an edge-brightened linear polarization near the core in the active jet state in 2016, similar to the quiescent jet state in 2008--2013. These observations indicate that the $\gamma$-ray flares may originate from compression of the knots by a standing shock in the core and the jet might consist of multiple complex layers showing time-dependent behavior, rather than of a simple structure of a fast jet spine and a slow jet sheath.


\end{abstract}



\section{Introduction \label{sect1}}

The Large Area Telescope (LAT) on board the \emph{Fermi} satellite \citep{Atwood2009} has revealed that blazars, active galactic nuclei (AGNs) having highly collimated and relativistic jets closely aligned with our line of sight \citep{UP1995, Blandford2018}, make up the largest fraction of observed $\gamma$-ray sources (e.g., \citealt{Acero2015, Ackermann2015}). It is commonly assumed that inverse Compton (IC) scattering of soft photons off relativistic electrons in the jets is responsible for the $\gamma$-ray emission\footnote{In addition to this leptonic model, there are also hadronic models for $\gamma$-ray emission in blazars (e.g., \citealt{Mannheim1993}, see also \citealt{Boettcher2012} and \citealt{Bottcher2013} for review of the leptonic and hadronic models).}. However, both the location of the $\gamma$-ray emission sites in AGN jets and the origin of the seed photons, which are upscattered in energy by the IC process, are still a matter of debate. The seed photons could be synchrotron photons from the same electrons that up-scatter the photons (synchrotron self-Compton, SSC; e.g., \citealt{Maraschi1992}) or photons from sources outside the jets (external Compton, EC) such as the accretion disk (e.g., \citealt{Dermer1992}), the broad line region (BLR, e.g., \citealt{Sikora1994}), and the dusty torus (DT, e.g., \citealt{Blazejowski2000}), or photons from the cosmic microwave background (e.g., \citealt{Tavecchio2000}).

Blazars can be divided into two classes based on their optical properties: flat spectrum radio quasars (FSRQs) and BL Lac objects (BL Lacs). This classification was initially phenomenological and based on the equivalent widths of emission lines being larger (FSRQs) or smaller (BL Lacs) than 5~\AA\ \citep{UP1995}. Eventually, it turned out that the different classes originated from different accretion regimes of AGNs, with FSRQs and BL Lacs having high and low accretion rates, respectively \citep{Ghisellini2011, HB2014, YN2014}. Their spectral energy distributions (SEDs) are distinct from each other. Compared to BL Lacs, FSRQ SEDs tend to show (i) higher luminosity, (ii) synchrotron and IC bumps peaking at lower observing frequencies, and (iii) a larger IC bump in comparison to the synchrotron one \citep{Fossati1998, Ghisellini1998, Ghisellini2017}. This behavior in FSRQs has been interpreted due to the efficient cooling of the relativistic electrons from the jets. The reason why the electrons cool so efficiently in FSRQs is thought to be because of the large amount of soft photons originating in the BLR. Since the BLR is thought to be within $10^3$--$10^4\ r_{\rm s}$ of the central engine (where $r_{\rm s}$ is the Schwarzschild radius), this is referred to as the ``near-dissipation zone'' scenario (e.g., \citealt{Ghisellini1998, Hartman2001, Ghisellini2010}).

However, many observations disfavor this scenario. For example, a significant fraction of $\gamma$-ray flares in blazars occur when superluminal knots in the jets pass through the radio core. The core is a (quasi-)stationary compact emission feature located at the upstream end of the jet (e.g., \citealt{Jorstad2001, JM2016}) resolved by very long baseline interferometry (VLBI). The core is often identified with a recollimation shock which may form when there is a pressure mismatch between the jet and the confining medium (e.g., \citealt{Sanders1983, WF1985, DM1988, Gomez1995, KF1997, Agudo2001, Cawthorne2013, Mizuno2015, Fromm2016, Fuentes2018, Park2018}) and is usually expected to be located quite far from the jet base, i.e., at distances $\gtrsim1$ pc in the source frame (e.g., \citealt{OG2009, Pushkarev2012}). This distance is larger than $10^4\ r_{\rm s}$ for most blazars and supports the ``far-dissipation zone'' scenario for the $\gamma$-ray flares. Likewise, the detection of very high energy (VHE, where VHE is defined as photon energies $>$100 GeV and high energy, HE, as $>$100 MeV) emission in several FSRQs (e.g., \citealt{Aleksic2011a, Aleksic2011b, Aleksic2014}) is challenging to explain with the near-dissipation zone scenario because it is difficult for the VHE photons to escape the intense radiation field of the BLR (e.g., \citealt{LB2006, TM2009, Barnacka2014}). On the other hand, it has been pointed out that the external seed photon field at the VLBI core would be too weak to produce the observed $\gamma$-ray emission (e.g., \citealt{Marscher2010, Aleksic2014}).

PKS 1510--089 is one of the brightest and most active blazars observed by \emph{Fermi-}LAT (e.g., \citealt{Abdo2010}) and has been detected at VHE bands \citep{HESS2013, Aleksic2014, Ahnen2017, MAGIC2018, Zacharias2019}. \cite{Marscher2010} detected a systematic rotation of the optical electric vector position angle (EVPA), followed by strong optical and $\gamma$-ray flaring that was also coincident with an ejection of a new superluminal knot from the core in 2009. They concluded that the $\gamma$-ray flares occurred in the superluminal knot as it passed through the core \citep{Marscher2008}. The origin of the seed photons was discussed in the context of a spine-sheath jet structure, where a relatively slow jet sheath surrounds a fast jet spine (see, e.g., Fig. 1 of \citealt{Ghisellini2005}, see also \citealt{Sol1989, Laing1996}). In contrast, based on (a) the absence of a correlation between X-ray and $\gamma$-ray fluxes in 2008 and 2009 and (b) a comparison of observed $\gamma$-ray-to-optical flux ratios to simulated ones, \cite{Abdo2010} concluded that the $\gamma$-ray emission is dominated by the EC process with the seed photons originating in the BLR. \cite{Dotson2015} suggested that some of the $\gamma$-ray flares in 2009 occurred at the distance of the DT, while others occurred in the vicinity of the radio core, by investigating the energy dependence of the flare decay time to infer the source of the seed photons.

\cite{Orienti2013} found a $\gamma$-ray flare from PKS 1510-089 in late 2011 after the onset of a strong radio flare and located the $\gamma$-ray emitting site to be about 10 pc downstream of the jet base. On the other hand, \cite{Saito2015} suggested that the $\gamma$-ray flares in 2011 occurred at the distance of 0.3--3 pc from the central engine with the seed photons provided by the BLR and DT, based on the model of internal shocks formed by colliding blobs of the jet plasma. \cite{Aleksic2014} showed the HE and VHE $\gamma$-ray spectra in 2012 smoothly connected with each other. The $\gamma$-ray light curves were correlated with the millimeter-wave light curves, and a superluminal knot emerged from the core near in time with the $\gamma$-ray flares. They showed that the observed SEDs could be explained well by two scenarios, (i) EC in the jet about 1 pc downstream of the central engine with seed photons from the DT and (ii) EC in the core at $\approx6.5$ pc downstream of the central engine with the seed photons being provided by the sheath. A recent study, using the Very Long Baseline Array (VLBA) at 43 GHz when the jet was in a quiescent state, revealed that the degree of linear polarization near the core increases toward the edges of the jet with the EVPAs predominantly perpendicular to the jet direction (\citealt{Macdonald2015}; see also \citealt{Macdonald2017} for the case of other blazars). This result indicates that there may be a relatively slow sheath of jet plasma surrounding the fast jet spine, as predicted in previous studies of $\gamma$-ray flares in this source (e.g., \citealt{Marscher2010, Aleksic2014}). The sheath could be an important source of seed photons in the far-dissipation zone scenario, and also can provide seed photons for ``orphan'' $\gamma$-ray flares that show little or no corresponding variability detected at longer wavelengths \citep{Macdonald2015}.

In 2015, PKS 1510--089 showed variable VHE emission on time scales of a few days during its long, elevated HE $\gamma$-ray state \citep{Ahnen2017}. This event was accompanied by a systematic optical EVPA rotation and the ejection of a knot from the core which was observed with the VLBA at 43 GHz, similar to the flares in 2009 \citep{Marscher2010} and 2012 \citep{Aleksic2014}. However, the knot (named K15) moved away from the core at a position angle (PA) radically different (by $\sim90^\circ$) from the historic jet direction \citep{Jorstad2017}. K15 was detected for five successive epochs from 2015 December to 2016 April and is unlikely to be an imaging artifact. \cite{Ahnen2017} could not determine if the ejection of this component is indeed related to the VHE or $\gamma$-ray emission in 2015 because of uncertainties in the kinematic analysis. 

The primary goal of this paper is to investigate the unusual kinematics and linear polarization structure of the jet in 2016 and 2017 and to probe a potential connection of the jet activity to the HE and VHE flares in 2015. Therefore, we extend the observational timeline of the kinematic analysis by \cite{Ahnen2017} by one year and four months. We refer the reader to other studies for detailed modeling of SEDs of our source with good spectral coverage in various periods (e.g., \citealt{Abdo2010, Dammando2011, HESS2013, Aleksic2014, Saito2015, Ahnen2017, MAGIC2018}).

This paper is organized as follows. We first present multi-wavelength light curves of PKS 1510--089 between 2013 and 2018 in Section~\ref{sect2}. In Section~\ref{sect3}, we focus on the peculiar behavior of the jet after the strong and complex multi-wavelength flare in 2015, by performing kinematic and linear polarization analysis. We discuss our results and draw our conclusions in Sections~\ref{sect4} and ~\ref{sect5}, respectively. In this paper, we adopt the following cosmological parameters: $H_0 = 70 {\rm\ km\ s^{-1}\ Mpc^{-1}}$, $\Omega_{\rm m} = 0.3$, and $\Omega_{\Lambda} = 0.7$, giving a projected scale of $5.0\ {\rm pc\ mas^{-1}}$ for PKS 1510-089 at a redshift of 0.36 \citep{Thompson1990}.

\section{Multi-wavelength Light Curves}
\label{sect2}

In this section, we present the long-term light curves of PKS 1510-089 at radio, optical, and $\gamma$-ray wavelengths, which is shown in Figure~\ref{lc}. We did not include X-ray light curves in our analysis because of relatively large time gaps in the \emph{Swift}-XRT light curve during the period of our interest; we refer the reader to \cite{MAGIC2018} for the long-term activity of our source at X-ray.

\subsection{iMOGABA}

The iMOGABA program observes about 30 $\gamma$-ray bright blazars with the Korean VLBI Network (KVN; \citealt{Lee2011, Lee2014}) at 22, 43, 86, and 129 GHz simultaneously (see \citealt{Lee2016} for details of the program). PKS 1510--089 has been observed almost every month since 2012 December. A standard data post-correlation process with the NRAO Astronomical Image Processing System (AIPS, \citealt{Greisen2003}) was performed by using the automatic pipeline for KVN data \citep{Hodgson2016}. We achieved high fringe detection rates and reliable imaging at up to 86 GHz by using the frequency phase transfer (FPT) technique \citep{Middelberg2005, Rioja2011, Rioja2014, Algaba2015, Zhao2018} which overcomes the rapid tropospheric phase variations characteristic for high frequencies. Nevertheless, the data at 129 GHz usually suffer from severe observing conditions such as relatively large sky opacity and low aperture efficiencies, which makes the detection rate lower than at other frequencies. Moreover, the 129 GHz results have larger uncertainties originating from inaccurate pointing and large gain errors (e.g., \citealt{Kim2017}). Thus, we excluded the 129 GHz data from our analysis. We used the Caltech \emph{Difmap} package for imaging and phase self-calibration \citep{Shepherd1997}. We performed a {\tt modelfit} analysis in \emph{Difmap} using circular Gaussian components. We note that we found a single component at the radio core in most epochs at all frequencies due to the compact source geometry and the relatively large beam size of the KVN. We generated radio light curves by using the flux density of the core component when a single component was detected and the total flux density when multiple components were detected (see the top panel of Figure~\ref{lc}).

\subsection{SMA}
\label{sectSMA}

\begin{figure}[!t]
\centering
\includegraphics[trim=12mm 6mm 7mm 8mm, clip, width = 0.49\textwidth]{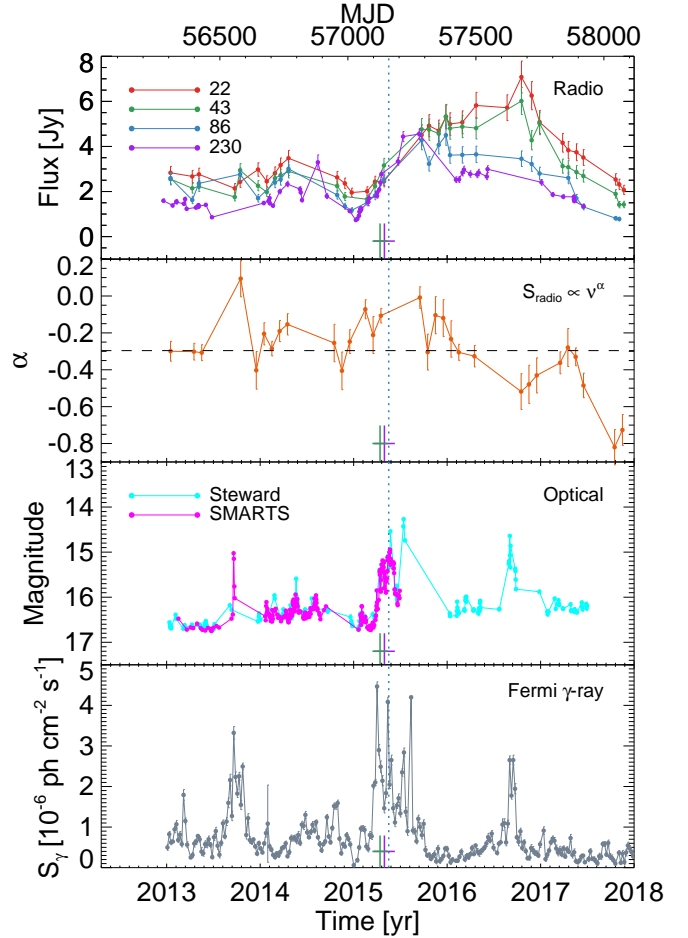}
\caption{\emph{Top panel:} Light curves of PKS 1510--089 from 2013 to 2018 at radio frequencies (22, 43, 86 GHz from the iMOGABA program, 230 GHz from the SMA). \emph{Second panel from top:} Spectral index obtained by fitting a simple power-law function to the radio spectra available for each time bin (see Section~\ref{sectspix}). \emph{Third panel from top:} Light curves at optical wavelengths (cyan: Steward observatory; magenta: SMARTS program). \emph{Bottom panel:} $\gamma$-ray light curve from \emph{Fermi}-LAT data. The crosses at the bottom show the epochs of zero separation of the knots K15 and J15 (vertical lines) with their $1\sigma$ errors (horizontal lines, see Section~\ref{sect3} and Figure~\ref{properties} for details). The blue vertical dotted line marks the time of VHE emission in mid-2015 \citep{Ahnen2017}. \label{lc}}
\end{figure}

The 230 GHz (1.3 mm) flux density data were obtained at the Submillimeter Array (SMA) near the summit of Mauna Kea (Hawaii). PKS 1510--089 is included in an ongoing monitoring program at the SMA to determine the fluxes of compact extragalactic radio sources that can be used as calibrators at mm wavelengths \citep{Gurwell2007}. Observations of available potential calibrators are from time to time observed for 3 to 5 minutes, and the measured source signal strength calibrated against known standards, typically solar system objects (Titan, Uranus, Neptune, or Callisto).  Data from this program are updated regularly and are available at the SMA website\footnote{\url{http://sma1.sma.hawaii.edu/callist/callist.html}} database \citep{Gurwell2007}. The light curve is shown in the top panel of Figure~\ref{lc}.

\subsection{Radio Spectral Index}\label{sectspix}

We obtained the radio spectral index as a function of time (the second panel from the top in Figure~\ref{lc}) by binning the light curves at 22, 43, 86, and 230 GHz into monthly time intervals. We then fitted the radio spectra with a simple power-law function, i.e., $S_\nu \propto \nu^\alpha$, for bins where flux data are available in at least three different frequency bands. One has to take into account synchrotron self-absorption to obtain more reliable fits to the radio spectra as done in other studies (e.g., \citealt{Fromm2011, Rani2013, Algaba2018}). However, we used simple power-law fitting in this work because of the limited spectral coverage in many time bins and because we could not find any significant deviation of the data from power-law fits within errors. The simple power-law fitting would be enough to show the long-term evolution of the radio spectral index, which fits our purpose. 


\subsection{Optical Photometric Data}

We collected publicly available optical photometric data from the Steward Observatory blazar monitoring program\footnote{\url{http://james.as.arizona.edu/~psmith/Fermi}} measured in the 500-to-700~nm band (see \citealt{Smith2009} for details) for the same period for which we obtained the \emph{Fermi} $\gamma$-ray data. We also obtained optical V band data from 2013 to mid-2015 from the Small and Moderate Aperture Research Telescope System (SMARTS\footnote{\url{http://www.astro.yale.edu/smarts/glast}}) monitoring program of \emph{Fermi} blazars (see \citealt{Bonning2012} for details). The optical light curves from the two datasets are shown in the second panel from the bottom in Figure~\ref{lc}.

\subsection{Fermi-LAT}

We followed \cite{MAGIC2018} for extracting the LAT $\gamma$-ray light curves. We used the \emph{Fermi}-LAT data observed in survey mode.\footnote{\url{https://fermi.gsfc.nasa.gov/cgi-bin/ssc/LAT/LATDataQuery.cgi}} We analyzed photons in the ``source event'' class using the standard ScienceTools (software version v11r5p3) and instrument response functions P8R2\_SOURCE\_V6 and the \emph{gll\_iem\_v06.fits} and \emph{iso\_P8R2\_SOURCE\_V6\_v06.txt} models for the Galactic and isotropic diffuse emission \citep{Acero2016}, respectively. We analyzed a region of interest (ROI) of $20^\circ$ radius centered at the position of PKS 1510--089. A zenith angle cut of $<90^\circ$ was applied to reduce contamination from the Earth's limb. We first performed an unbinned likelihood analysis using gtlike \citep{Acero2015} for the events recorded from 2013 February 1 to June 30 (MJD 56324--56474) in the energy range between 100 MeV and 300 GeV. The model parameters for the sources within $10^\circ$ of the center of the ROI were left free, while the parameters for the sources from $10^\circ$ to $20^\circ$ were fixed to their 3FGL catalog values for this first unbinned likelihood analysis. \cite{MAGIC2018} found no new strong sources within $20^\circ$ of PKS 1510--089 in other time ranges; we conclude that the best-fit parameters obtained from the unbinned likelihood analysis for the five-month period we studied are representative for other periods of interest. For further analysis, we removed sources with a test statistic (TS; \citealt{Mattox1996}) less than 9, corresponding to $\approx3\sigma$ detections. We then generated a light curve binned to one-week time intervals of PKS 1510--089 at $E>100$ MeV by fixing the model parameters for all the sources using the output model in the first unbinned likelihood analysis, except for our target and the variable sources reported in the 3FGL catalog \citep{Acero2015}. We fitted a power law spectrum with both the flux normalization and the spectral index being free parameters for these sources. We note that the normalization of the Galactic and isotropic diffuse emission models were also left free. The $\gamma$-ray light curve is shown in the bottom panel of Figure~\ref{lc}.

\begin{figure*}[!t]
\centering
\includegraphics[trim=3mm 2mm 5mm 3mm, clip, width = \textwidth]{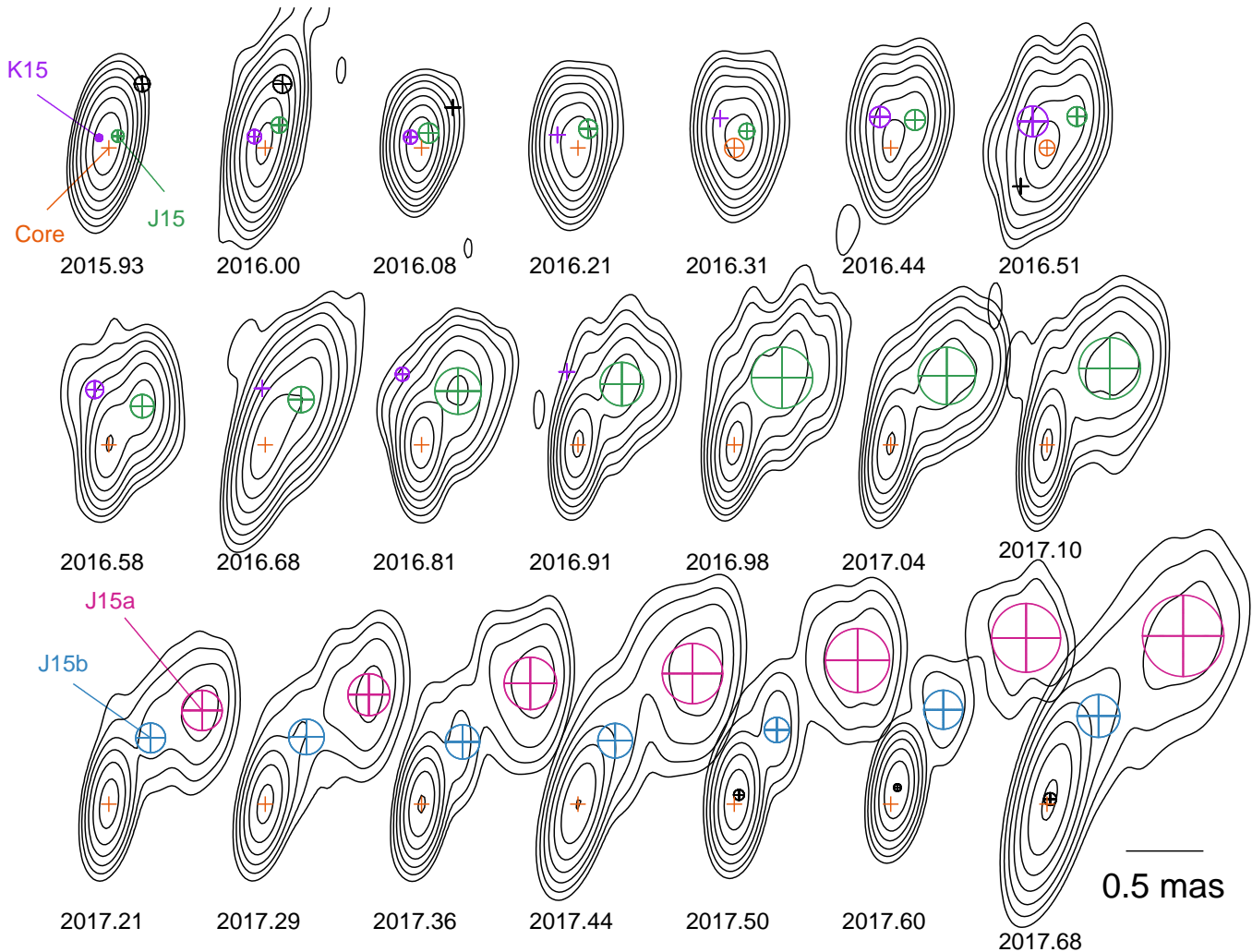}
\caption{A series of CLEAN maps of PKS 1510--089 obtained by the VLBA 43 GHz data. Contours start from 25 mJy/beam and increase by factors of two. Circular Gaussian {\tt modelfit} components are shown as crosses surrounded by circles overlaid on the contours. Crosses without surrounding circles show components with sizes smaller than 0.04 mas, corresponding to $\approx1/5$ of the synthesized beam size. Components of the same color in different epochs are identified as being the same object. Black components are not used for component identification. The epoch of observation of each map in decimal years is noted below the contours. The dark solid line in the bottom right corner illustrates the angular scale in the images. \label{clean}}
\end{figure*}

\begin{figure}[!t]
\centering
\includegraphics[trim=10mm 5mm 4mm 5mm, clip, width = 88mm]{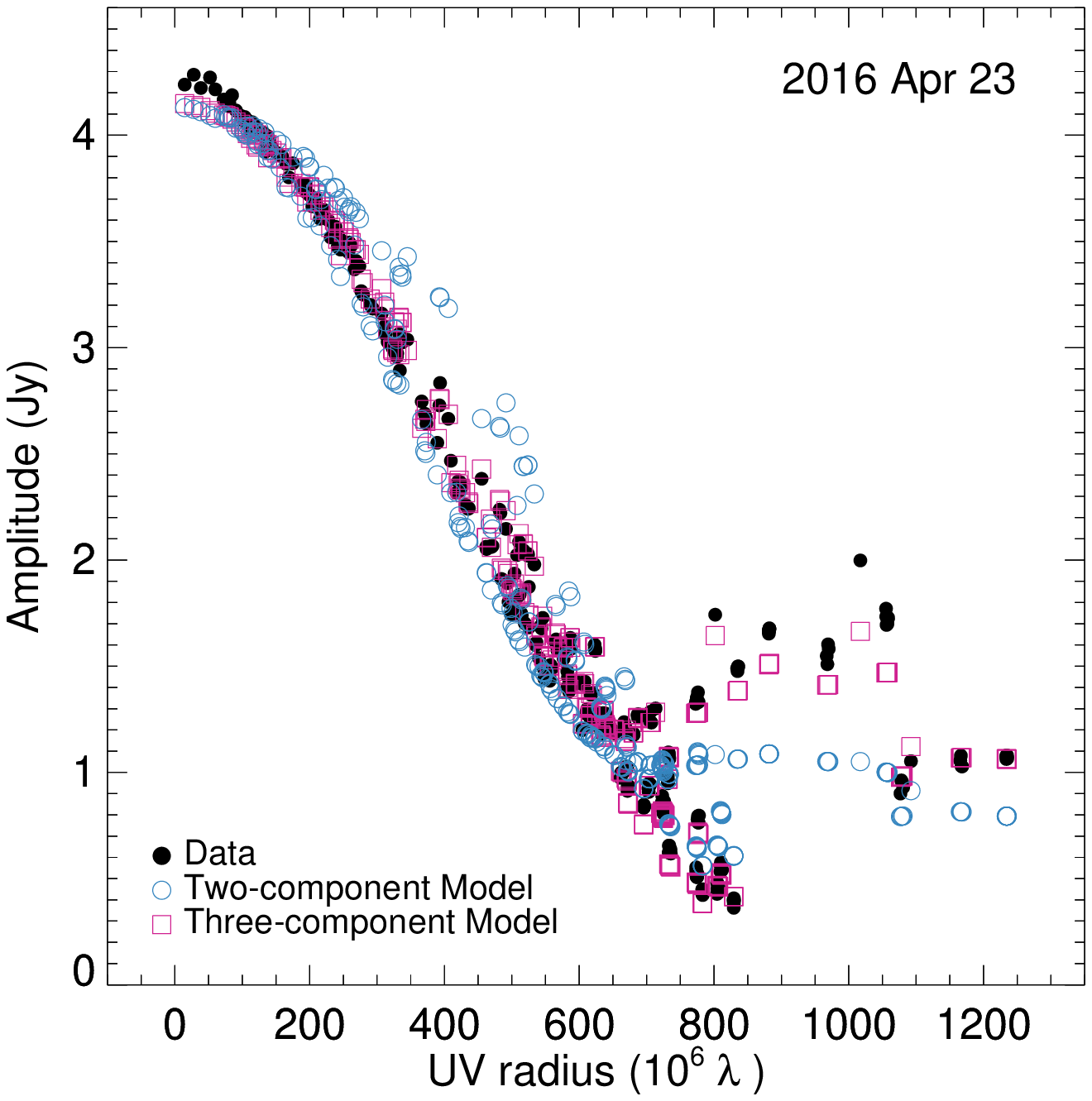}
\caption{Visibility amplitude as a function of $uv$-distance. The VLBA 43 GHz data observed on 2016 Apr 23 are shown with the black data points, while the models fitted with two and three circular Gaussian components are shown with the blue and red data points, respectively. The model with three Gaussian components fits the data better than that with two Gaussian components (see Section~\ref{sect3} for details). \label{radpl}}
\end{figure}

\section{Jet kinematics and linear polarization analysis}
\label{sect3}

We used the calibrated VLBA data observed over 21 epochs from 2015 December to 2017 September taken from the VLBA-BU-BLAZAR program\footnote{\url{https://www.bu.edu/blazars/VLBAproject.html}} except for 2016 October 6 because two antennas were unable to observe at that time. The details of the observations and the data reduction are described in \cite{Jorstad2005, Jorstad2017}. We performed a {\tt modelfit} analysis of the visibility data in \emph{Difmap} for each epoch using multiple circular Gaussian components. We present the {\tt modelfit} components overlaid on the CLEAN images in Figure~\ref{clean}. We first identified the radio core as the compact and bright component located at the upstream end of the jet. We assumed that the core is stationary and identified its location with the origin of each map; we then identified other components (shown in the same color in different epochs in Figure~\ref{clean}). A triple-component structure is consistently found in the first 11 epochs. The two jet components labeled K15 (following \citealt{Ahnen2017}) and J15, are moving away from the core. K15 fades out and is no longer detected after 2016.91, while J15 is continuously moving with an average PA (measured north through east with respect to the core) of $-37^\circ$. Since 2017.21, J15 appears to have split into two components, labeled J15a and J15b.

\begin{figure}[!t]
\centering
\includegraphics[trim=5mm 1mm 4mm 3mm, clip, width = 88mm]{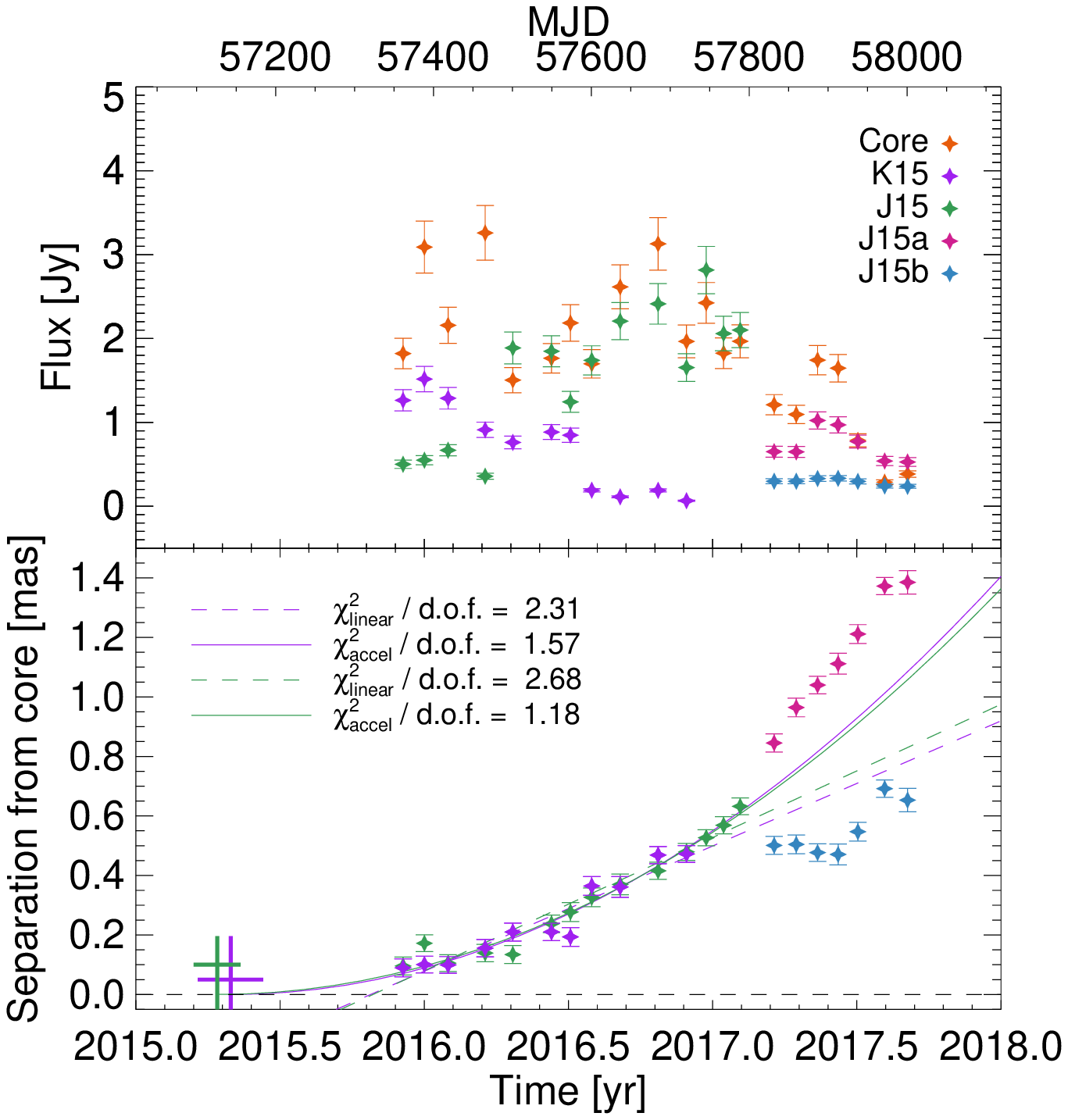}
\caption{Flux density (top) and separation from core (bottom) as functions of time for all identified components (with the same color coding as in Figure~\ref{clean}). The solid curves and the dashed lines in the bottom panel (purple for K15 and green for J15) are the best-fit curves assuming acceleration and constant velocity, respectively. The reduced $\chi^2$ ($\chi^2 / d.o.f.$, where $d.o.f.$ denotes the degree of freedom) values are noted for each best-fit function. The crosses in the bottom left corner show the  zero-separation epochs (vertical lines) with their $1\sigma$ errors (horizontal lines). \label{properties}}
\end{figure}

Interestingly, both K15 and J15 are seen in the five epochs of the VLBA 43 GHz data from 2015.93 to 2016.31 presented in \cite{Ahnen2017} -- however, they identified J15 as the core probably because the distance between the core and J15 is quite small, $\approx0.1$ mas, in these epochs. \cite{Casadio2017} presented a map obtained using the global millimeter VLBI array (GMVA) at 86 GHz in 2016 May and found a compact triple component structure within the central $\sim0.5$ mas. Their results motivated us to fit models with three Gaussian components near the core to the data and we found that they provide us with better fits in terms of reduced $\chi^2$ in all five epochs. Specifically, the reduced $\chi^2$ is 0.5, 5.6, 0.7, 4.1, and 2.8 in the {\tt modelfit} results we present here\footnote{We excluded the components outside the region around the core, e.g., at distances larger than 0.3 mas from the map center, to ensure a proper comparison.} and is 2.0, 26.6, 2.7, 9.7, and 20.9 when using two Gaussian components, respectively, in chronological order. To demonstrate how the three components improve the goodness of fit, we selected an epoch within a month of the GMVA observation and present the visibility amplitudes of the data as a function of $uv$-radius (black data points in Figure~\ref{radpl}). We can see that the model with three Gaussian components (red) describe the observed data better than the model with two components (blue) at various $uv$-radius. In addition, K15 and J15 are also consistently seen in later epochs, making it highly unlikely that they are artifacts. While the PA of J15 ($\approx-37^\circ$ on average) seems to be generally consistent with the global jet direction on the same spatial scale (PA of $\approx-34^\circ$, \citealt{Jorstad2017}), the PA of K15 ($\approx+28^\circ$ on average) is significantly different.

\begin{figure*}[!t]
\centering
\includegraphics[trim=7mm 1mm 2mm 2mm, clip, width = 0.67\textwidth]{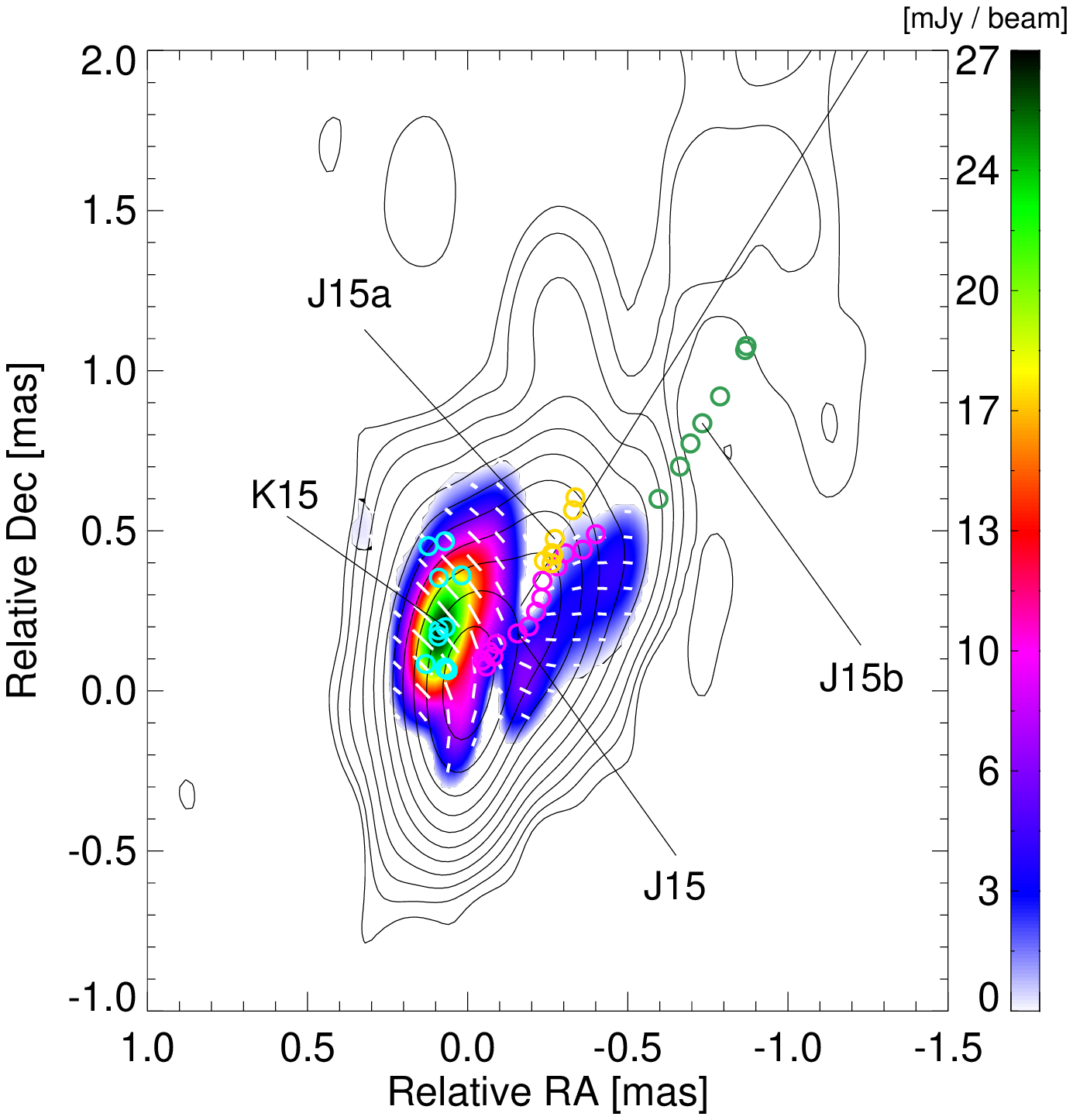}
\caption{Stacked linear polarization map of the data observed in 11 epochs between 2015.93 and 2016.91 (see Section~\ref{sect3} for details). Color shows linear polarization intensity; contours show total intensity. A color scale is shown in the vertical bar on the right side. White ticks show EVPAs. We present the positions of identified components in different epochs with open circles in the map (cyan for K15, magenta for J15, yellow for J15a, and green for J15b). The black solid line indicates the global jet direction on mas scales, with a PA of $-34^\circ$ \citep{Jorstad2017}. \label{stack}}
\end{figure*}

We present the flux density and the separation from the core as functions of time for different components in Figure~\ref{properties}. The light curves for each component show moderate variability but are continuous across multiple epochs in general, suggesting that the identification of components with specific jet regions is reliable. As for the separation from the core, we fitted both linear functions (i.e., motions with constant velocities) and parabolic functions (i.e., accelerated motions) to K15 and J15 and found that the latter provides us with better fits in terms of reduced $\chi^2$ (see Figure~\ref{properties}). The separation from the core for J15a, which might be the same knot as J15 but cannot be tested straightforwardly, is in a better agreement with the acceleration motion of J15 than with the linear motion. The zero-separation epochs, i.e., the time when the components are expected to emerge from the core, are $2015.33\pm0.11$ and $2015.28\pm0.08$ for K15 and J15 (corresponding to MJD $57144\pm42$, $57127\pm30$), respectively. These estimates are slightly earlier than the ones by \cite{Ahnen2017}, MJD $57230\pm52$, by 1--2$\sigma$, presumably because of different component identification and a smaller number of data points in their work.

We further checked if the two distinct emission regions can also be detected in linear polarization maps. Following \cite{Macdonald2015}, we generated a stacked polarization image by (i) convolving maps for all Stokes parameters from different epochs with the same beam (average full width at half maximum with major axis, minor axis, and PA of 0.41 mas, 0.15 mas, $-6.62^\circ$, respectively), (ii) aligning the maps such that the radio core is at the origin, and (iii) averaging the maps for each Stokes parameter. We used the epochs between 2015.93 and 2016.91, for which we could find both K15 and J15 in the total intensity maps. The results are presented in Figure~\ref{stack}. We note that we did not take into account Faraday rotation for our further analysis because the Faraday rotation measure was observed to be $165\rm\ rad/m^2$ (corresponding to EVPA rotation by $\lesssim1^\circ$ with respect to the intrinsic EVPA at 43 GHz) at the 15 GHz core \citep{Hovatta2012} and K15 and J15 are most likely located downstream of the 15 GHz core. We found that significant polarized emission is detected in the regions corresponding to K15 and J15. The eastern polarization component shows relatively strong and compact polarized emission with EVPAs almost perpendicular to the jet axis, while the western component shows polarized emission extended along the direction close to the global jet direction with EVPAs oblique to the jet axis.

\section{Discussion}
\label{sect4}

\subsection{Comparison of the $\gamma$-ray flares in 2015 with previous flares \label{sect41}}

In 2015, PKS 1510--089 was in an active $\gamma$-ray state which lasted for more than six months (Figure~\ref{lc}, see also \citealt{Ahnen2017, Prince2017, MAGIC2018}). Optical flares also occurred at about the same time as the $\gamma$-ray flares, while a strong radio flare lasting $\gtrsim2$ years started in 2015. The 37-GHz radio light curve presented in a recent study indicates that the radio flare consists of two separate flares, one starting near MJD 57000 and the other near MJD 57600 \citep{MAGIC2018}. The latter seems to be related to the $\gamma$-ray and optical flares in mid-2016. Variable VHE emission was detected by the MAGIC telescopes on MJD 57160 and 57165. The VHE radiation seems to originate from the same region that emitted the HE $\gamma$-ray and optical flares \citep{Ahnen2017}. This is consistent with the time when K15 and J15 emerged from the core, which suggests that these components may be responsible for the multi-wavelength flares, including the VHE emission, in mid-2015.

The flares in 2015 are remarkably similar to the ones in 2009 \citep{Marscher2010} in the sense that (i) the $\gamma$-ray flares are nearly simultaneous with the optical flares, (ii) a systematic rotation of EVPAs at optical wavelengths is detected \citep{Ahnen2017}, (iii) new jet components emerge from the core during the flares, and (iv) VHE emission is detected \citep{HESS2013}. Therefore, a similar interpretation based on the far-dissipation zone scenario, compression of the knots by a standing conical shock in the core leading to strong $\gamma$-ray flares \citep{Marscher2008, Marscher2010, Marscher2014}, can be applied to the 2015 flares. Indeed, the radio light curves show optically thick spectra when the emerging knots are close to the core (with $\alpha$ in the range from $-0.3$ to $0$ as seen in Figure~\ref{lc}), while they become optically thin after the knots are well separated from the core in later epochs ($\alpha$ from $-0.8$ to $-0.3$). This behavior is in good agreement with the prediction of the shock-in-jet model (e.g., \citealt{MG1985, Valtaoja1992, Fromm2011, Hughes2011}), supporting the above interpretation. However, there is a remarkable difference in the behavior of the jet: we found two laterally separated moving knots emerging nearly simultaneously from the core, whereas a single knot was detected in 2009 \citep{Marscher2010}.

\subsection{Double-knot Jet Structure\label{sect42}}

Blazars usually display a ridge-brightened, knotty jet structure\footnote{We note, however, that some blazars show rapid changes in apparent jet position angles in projection on the sky plane, which might be related to radio flares and $\gamma$-ray flares in those sources (e.g., \citealt{Agudo2012, Hodgson2017}).} (e.g., \citealt{Jorstad2005, Jorstad2017, Lister2013, Lister2016}) and the double knot structure seen in PKS 1510--089 is uncommon. We found that the linear polarization structure near the core is characterized as two distinct polarized regions (see Figure~\ref{stack}). Interestingly, this behavior was observed in the active state, when the two knots (K15 and J15) emerge from the core and propagate outwards during the period of 2015.93 -- 2016.91, while a rather similar linear polarization structure was observed in the quiescent jet state in 2008 -- 2013 also \citep{Macdonald2015}. We present the positions of identified {\tt modelfit} components (see Section~\ref{sect3} and Figure~\ref{clean}) on top of the stacked polarization map. The compact polarized emission on the east side of the core is overlapped with the positions of K15, while the extended polarized emission on the west side of the core is distributed along with the trajectory of J15. Thus, the eastern and western polarized emission seems to be associated with the moving knots K15 and J15, respectively.

One of the possible origins of the double knot structure and the corresponding linear polarization structure is a large-scale helical magnetic field permeating in the jet (e.g., \citealt{Lyutikov2005, Clausen-brown2011, Murphy2013, Zamaninasab2013}). Evidence for helical magnetic fields in the jets of at least some blazars was provided by VLBI observations of Faraday rotation in the jets (e.g., \citealt{Asada2002, Algaba2013, Zamaninasab2013, Gomez2016, Gabuzda2018}). The helical field, depending on the jet viewing angle and the field pitch angle, can produce asymmetric profiles of both total intensity emission and linear polarization emission transverse to the jet. However, the transverse total intensity profile for blazars is expected to be more or less symmetric (see the case of $\theta_{\rm ob}\Gamma=1/1.2$ or 1/2 in Figure 2 in \citealt{Clausen-brown2011}). PKS 1510--089 is a highly beamed blazar for which $\theta_{\rm ob}\Gamma=0.47-1.23$ is expected ($\theta_{\rm ob}=1.2-3.4^\circ$ and $\Gamma=20.6-36.6$; \citealt{Jorstad2005, Jorstad2017, Hovatta2009, Savolainen2010}), and the observed complicated evolution of the total intensity profile, characterized by gradually decreasing and increasing flux densities of K15 and J15 over time, respectively (Figure~\ref{properties}), would be difficult to explain with the helical field scenario.


Another possible explanation is a spine-sheath structure in the jet, with a relatively slow sheath of jet plasma surrounding the fast jet spine. Such a structure is suggested by the \emph{limb brightening} of the jets observed in several sources (e.g., \citealt{Giroletti2004, Nagai2014, Hada2017a, Giovannini2018}) and was also introduced in theoretical modeling to explain the discrepancy between high Doppler factors\footnote{$\delta = 1/\Gamma(1-\beta\cos\theta_{\rm ob})$ with $\Gamma$, $\beta$, and $\theta_{\rm ob}$ being the jet bulk Lorentz factor, intrinsic velocity, and the viewing angle, respectively} necessary to explain the TeV-detected BL Lacs and FR I \citep{FR1974} radio galaxies and the rather slow jet motions observed in those sources (see e.g., \citealt{Ghisellini2005}, see also \citealt{TG2008, TG2014}). One of the observational signatures of a spine-sheath structure is an orientation of EVPAs perpendicular to the jet axis in the sheath (e.g., \citealt{Attridge1999, Pushkarev2005}). The sheath is thought to be generated by shear between the relativistic jet plasma and the ambient medium. At the boundary, the plasma jet and the embedded helical or tangled magnetic field are stretched along the direction of propagation of the jet due to the velocity gradients between the two layers \citep{Wardle1994}. This leads to an increase in the fractional polarization towards the jet edges, with the magnetic field being predominantly parallel to the jet direction, and thus EVPAs being perpendicular to the jet direction for an optically thin jet \citep{Pacholzcyzk1970}. Remarkably, \cite{Macdonald2015} suggested that the edge-brightened linear polarization structure near the core observed in the quiescent jet state in 2008 -- 2013 is consistent with the presence of a jet sheath, which can be an important source of seed photons for the orphan $\gamma$-ray flare observed in this source in 2009.

At a first glance, the observed features of K15, i.e., (i) a significant offset of PA from the global jet direction, and (ii) significant polarized emission with EVPAs perpendicular to the jet direction, are reminiscent of the sheath\footnote{In this scenario, a possible reason for the sheath appearing on only one side of the jet (K15) is that the interaction of the jet with the ambient medium is strongest on this side. The trajectory of J15 (Figure~\ref{stack}) follows the jet axis in the first five epochs but then shows a slightly curved trajectory towards the opposite side to K15, supporting this conjecture (see \citealt{Attridge1999} for a similar case observed in 1055+018).} on the east side of the core detected in the quiescent jet state \citep{Macdonald2015}. On the other hand, those of J15, i.e., (i) a trajectory in agreement with the global jet direction, and (ii) the extended polarized emission region along its trajectory with EVPAs oblique to the jet axis, are in agreement with a jet spine which is possibly a propagating shock (e.g., \citealt{Hughes2005, Jorstad2007}). However, our kinematic results suggest that both knots are moving at similar apparent speeds (Figure~\ref{properties}), which is not consistent with the scenario that K15 and J15 are a slow jet sheath and a fast jet spine, respectively. This indicates that a simple spine-sheath scenario may not be able to explain the observed kinematics of these knots.

\vspace{0.39cm}

\subsection{Acceleration motions and Spine-sheath Scenario \label{sect43}}

The apparent motions of K15 and J15 gradually accelerate from $\approx5c$ to $\approx13c$, with $c$ being the speed of light (corresponding to $\Gamma$ from $\approx11$ to $\approx19$ for $\theta_{\rm ob} = 1.2^\circ$ and from $\approx7$ to $\approx13$ for $\theta_{\rm ob} = 3.4^\circ$ for the fixed viewing angles), and possibly up to $\approx28c$ for J15 if it can be identified with J15a in later epochs. The observed acceleration of apparent speeds of these knots could be due to a change of the viewing angles, or the bulk Lorentz factors, or both. On the one hand, the acceleration is observed within the physical, de-projected distance (from the core) of $\approx75$ pc, corresponding to $\lesssim 3\times10^6\ r_{\rm s}$, when using a jet viewing angle of $2.3^\circ$, an average of $1.2^\circ$ \citep{Jorstad2017} and $3.4^\circ$ \citep{Hovatta2009, Savolainen2010}, and a black hole mass of $M_{\rm BH} \approx 2.5\times10^8\ M_{\odot}$ \citep{PT2017}. This is beyond the scale of a so-called an acceleration and collimation zone, where AGN jets are expected to be substantially collimated and accelerated to relativistic speeds through a magnetohydrodynamic process (e.g., \citealt{Meier2001, VK2004, Komissarov2007, Komissarov2009, Tchekhovskoy2008, Lyubarsky2009}). This process is believed to occur within the distances of $\lesssim10^4-10^6\ r_{\rm s}$ from the jet base (e.g., \citealt{Marscher2008}), and has been observed for the nearby radio galaxies M87 and Cygnus A \citep{AN2012, Asada2014, Boccardi2016, Mertens2016, Hada2017b, Walker2018}. On the other hand, bulk jet acceleration of blazars within deprojected distances of $\approx100$ pc from the core was found to be common (e.g., \citealt{Homan2015}) and the exact scale of the acceleration and collimation zone of blazars is under debate (e.g., \citealt{Hada2018}). Thus, we could not exclude the possibility that the observed acceleration of K15 and J15 is due to a change in the Lorentz factors.


\begin{figure}[!t]
\centering
\includegraphics[trim=7mm 10mm 2mm 5mm, clip, width = 88mm]{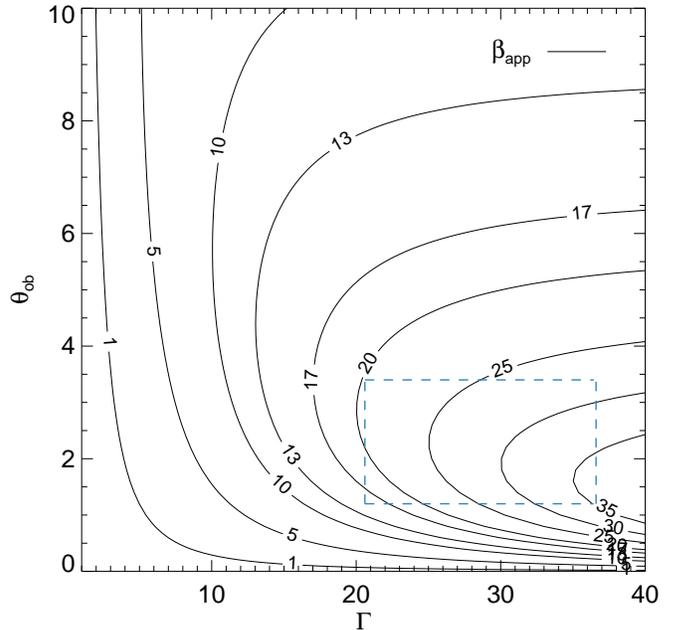}
\caption{Contours show apparent speed in units of the speed of light on a $\Gamma$ -- $\theta_{\rm ob}$ plane with the value for each contour noted. The blue rectangle shows the expected range of $\Gamma$ and $\theta_{\rm ob}$ based on previous studies \citep{Jorstad2005, Jorstad2017, Hovatta2009, Savolainen2010}. \label{doppler}}
\end{figure}

Given that the PA and linear polarization properties of K15 and J15 could be consistent with a spine-sheath structure but their apparent motions are not necessarily consistent with this scenario (Section~\ref{sect42}), we considered the possibility that J15 is \emph{intrinsically} much faster than K15, i.e., $\Gamma_{\rm J15}\gg\Gamma_{\rm K15}$, but that both components show similar \emph{apparent} motions due to different viewing angles. In Figure~\ref{doppler}, we present a contour plot of apparent speed on a $\Gamma$--$\theta_{\rm ob}$ plane. Assuming that the acceleration of the apparent speeds of K15 and J15 is purely due to a gradual increase in the viewing angle, $\Gamma_{\rm K15}\gtrsim13$ is necessary. Even if we assume that the intrinsic speed of J15 is very fast with $\Gamma_{\rm J15}=30$ based on previous studies \citep{Jorstad2005, Jorstad2017, Hovatta2009, Savolainen2010, Lister2016}, the ratio of $\Gamma_{\rm J15}$ to $\Gamma_{\rm K15}$ would be only a few, while the spine-sheath model usually assumes the ratio of $\Gamma_{\rm spine}$ to $\Gamma_{\rm sheath}$ to be $\approx10$ to explain the observed SEDs (e.g., \citealt{Ghisellini2005, Aleksic2014, Macdonald2015}). We note that the results of jet kinematics could also change if the absolute position of the core is changing during the radio flare (e.g., \citealt{Lisakov2017, Hodgson2017}), though we could not find any obvious systematic shifts in the positions of the downstream jet components. If a similar double knot structure is detected in the future, phase-referencing observations would help constrain the velocities of those knots more accurately (e.g., \citealt{Niinuma2015}).

We note that similar laterally extended jet emission, so-called off-axis jet emission, was observed in other $\gamma$-ray bright blazars such as 3C~279 \citep{Lu2013} and Mrk~501 \citep{Koyama2016}. \cite{Koyama2016} proposed two possible explanations for the off-axis jet emission. It could be either (i) an internal shock formed on an axis different from the global jet axis, or (ii) a part of a dim and slow outer layer that is Doppler boosted at the time of observation. The latter scenario is in agreement with the spine-sheath scenario. We note that, in any case, the off-axis emission (corresponding to K15 in our case) must be significantly Doppler-boosted at the time of observations with a similar Doppler factor to that of the main jet, otherwise, we would not observe the off-axis emission unless its synchrotron emissivity was much higher than the main jet emission, which we consider unlikely. The jet has previously only shown knots along the global jet direction (e.g., \citealt{Lister2016, Jorstad2017}), yet has shown evidence for a layered structure from linear polarization observations \citep{Macdonald2015}. Taken as a whole, we conjecture a scenario that the jet emission from the off-axis layer, persistently existing in this source, would be visible only when it is significantly Doppler-boosted, which was realized after the strong optical and $\gamma$-ray flares in 2015 and during the strong radio flare in 2016 (Figure~\ref{lc}).


\subsection{Origin of the 2015 $\gamma$-ray flare}

Our jet kinematic results imply that the strong HE and VHE flares in 2015 could be related to the ejection of K15 and J15 from the core (Figures~\ref{lc} and~\ref{properties}). As already noted in Section~\ref{sect41}, the ejection of new knots coincident with $\gamma$-ray flares was observed many times during previous HE and VHE flares in this source, which led previous studies to conclude that the core might be a dominant emission site of those flares (e.g., \citealt{Marscher2010, Orienti2013, Aleksic2014}). The location of the core at 43 GHz, as derived from a core-shift analysis \citep{Pushkarev2012}, is 5.3 -- 15.0 pc downstream of the jet apex, depending on the assumed jet viewing angle. This is too distant for the DT to provide the relativistic electrons in the core with enough seed photons \citep{Marscher2010, Aleksic2014}. Accordingly, additional seed photons from a slower sheath surrounding the jet spine, which may not be detected in usual cases due to small Doppler boosting, have been considered. This could explain (i) the highly variable $\gamma$-ray-to-optical flux ratio for different flares during the active $\gamma$-ray state in 2009 \citep{Marscher2010}, (ii) the SEDs, including the VHE emission, observed in 2012 \citep{Aleksic2014}, and (iii) the orphan $\gamma$-ray flare in 2009 \citep{Macdonald2015}.

\begin{figure}[!t]
\centering
\includegraphics[trim=10mm 4mm 3mm 5mm, clip, width = 88mm]{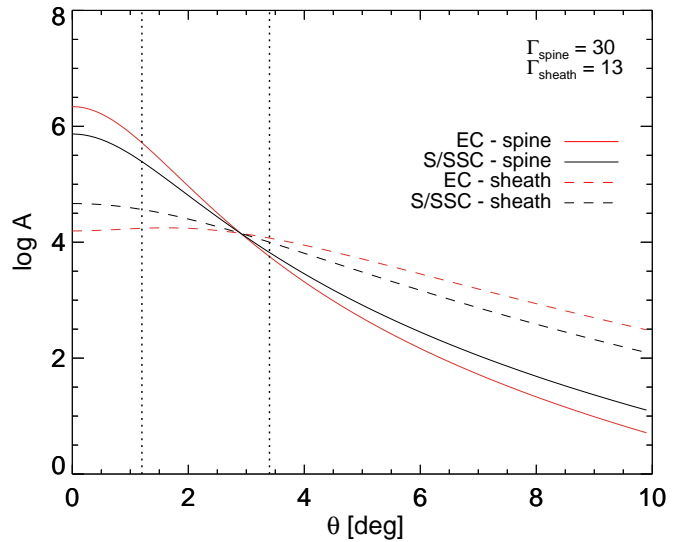}
\caption{Logarithmic amplification factors for different emission components as functions of jet viewing angle in the spine-sheath model with $\Gamma_{\rm spine}=30$ and $\Gamma_{\rm sheath}=13$ \citep{Ghisellini2005}. The solid (dashed) lines are for the spine (sheath) emission and the red (black) lines are for EC (S/SSC) emission. (S/SSC refers to ``synchrotron and SSC''.) The vertical dotted lines denote the jet viewing angles estimated in the literature \citep{Hovatta2009, Savolainen2010, Jorstad2017}. \label{amp}}
\end{figure}

However, our results show that the situation might be more complicated for the 2015 flares. In the spine-sheath model \citep{Ghisellini2005}, the EC intensity from the spine is amplified by a factor of $\delta_{\rm spine}^{3-\alpha}(\delta_{\rm spine}/\delta_{\rm sheath})^{1-\alpha}$, where $\delta_{\rm spine}$ and $\delta_{\rm sheath}$ are the Doppler factors of spine and sheath, respectively. The amplification factor for the synchrotron or SSC intensity from the spine is $\delta_{\rm spine}^{3-\alpha}$. The amplified sheath intensity is found analogously, i.e., by replacing $\delta_{\rm spine}$ by $\delta_{\rm sheath}$ and vice versa (see \citealt{Ghisellini2005} and \citealt{TG2008} for details). In Figure~\ref{amp}, we present the logarithmic amplification factors for synchrotron/SSC and EC emission in the spine and the sheath as functions of jet viewing angle with the assumed $\Gamma_{\rm spine}=30$ and $\Gamma_{\rm sheath}=13$ according to our consideration of K15 and J15 being a relatively slow jet sheath and a fast jet spine, respectively, in Section~\ref{sect43}. We used the average spectral index of $\alpha=-0.3$ obtained in Section~\ref{sectspix}. The ratio of the amplification factors of EC and synchrotron radiation of the spine is less than $\approx2$ for the expected viewing angle range for PKS 1510--089, while the observed peak luminosity of the IC component in 2015 is more than an order of magnitude larger than that of the synchrotron component \citep{Ahnen2017}.

Therefore, the ejection of double knots from the core near the time of the HE and VHE flares in 2015 supports the ''far-dissipation zone'' scenario with the core being a dominant emission site of $\gamma$-ray flares, while the observed motions of the knots make it difficult to reconcile with a spine-sheath jet structure needed for this scenario. One possible explanation is that the sheath itself may consist of multiple layers showing time-dependent behavior. What we observed as off-axis jet emission could be a layer with relatively fast speed. 

An alternative scenario as suggested by \cite{Ahnen2017} places the $\gamma$-ray emission region at $\approx0.2$ pc from the central engine. In this scenario, most of the seed photons for the external Compton processes would be provided by the DT. This one-zone model could successfully describe the observed SEDs including the VHE emission in 2015. However, the core-shift analysis of \cite{Pushkarev2012} placed the location of the 43 GHz core to be at $\sim10$ pc from the jet base. If the assumptions in their core-shift study are correct, this would suggest that the kinematic association with the $\gamma$-ray flaring is coincidental. A possibility to reconcile these results could be that the assumptions underlying the core-shift analysis such as the equipartition between jet particles and magnetic field energy densities and a smooth radially expanding jet may not hold. Additionaly, the core-shift can be time-dependent \citep{Niinuma2015}, potentially explaining the discrepancy.

\section{Conclusions}
\label{sect5}

In 2015, PKS 1510--089 showed an active $\gamma$-ray state observed by \emph{Fermi}-LAT with variable VHE emission detected by the MAGIC telescopes. We performed a jet kinematic analysis using VLBA 43 GHz data observed in 21 epochs between late 2015 and mid-2017. We found that two laterally separated knots in the jet nearly simultaneously emerge from the radio core during the period of $\gamma$-ray flaring and VHE emission in 2015. From the KVN and SMA monitoring data, we found that the onset of a strong multi-band radio flare begins near in time with the $\gamma$-ray flares, showing an optically thick spectrum at the beginning and gradually becoming optically thin as the knots become well separated from the core. Likewise, multiple complex optical flares and a systematic EVPA rotation occur along with the $\gamma$-ray flares \citep{Ahnen2017}. These observations suggest that the compression of moving knots by a standing conical shock in the core might be responsible for the HE and VHE flares. If the kinematic behavior is associated with the flaring, core-shift analysis indicates that the $\gamma$-ray emission region is $\sim10$ pc downstream of the jet base, which would supports the ``far-dissipation zone'' scenario. We found that many of the observed properties of the double knots are consistent with a spine-sheath jet structure, which has been invoked to resolve the problem of the lack of seed photons for external Compton processes in the far-dissipation zone scenario. However, the observed speeds of the knots are difficult to explain with the fast jet spine and slow jet sheath model, indicating that the jet may consist of multiple, complex layers with different speeds which themselves could be time-dependent.

\acknowledgments 
We thank the referee for constructive comments, which helped to improve the paper significantly. J.P. and J.-Y.K. thank Nicholas R. MacDonald for valuable discussions. We are grateful to the staff of the KVN who helped to operate the array and to correlate the data. The KVN and a high-performance computing cluster are facilities operated by the KASI (Korea Astronomy and Space Science Institute). The KVN observations and correlations are supported through the high-speed network connections among the KVN sites provided by the KREONET (Korea Research Environment Open NETwork), which is managed and operated by the KISTI (Korea Institute of Science and Technology Information). The Submillimeter Array is a joint project between the Smithsonian Astrophysical Observatory and the Academia Sinica Institute of Astronomy and Astrophysics and is funded by the Smithsonian Institution and the Academia Sinica. Data from the Steward Observatory spectropolarimetric monitoring project were used. This program is supported by Fermi Guest Investigator grants NNX08AW56G, NNX09AU10G, NNX12AO93G, and NNX15AU81G. This paper has made use of up-to-date SMARTS optical/near-infrared light curves that are available at \url{www.astro.yale.edu/smarts/glast/home.php}. This study makes use of 43 GHz VLBA data from the VLBA-BU Blazar Monitoring Program (VLBA-BU-BLAZAR; http://www.bu.edu/blazars/VLBAproject.html), funded by NASA through the Fermi Guest Investigator Program. The VLBA is an instrument of the National Radio Astronomy Observatory. The National Radio Astronomy Observatory is a facility of the National Science Foundation operated by Associated Universities, Inc. We acknowledge financial support from the Korean National Research Foundation (NRF) via Global Ph.D. Fellowship Grant 2014H1A2A1018695 (J.P.) and Basic Research Grant NRF-2015R1D1A1A01056807 (S.T., D.-W.K., J.C.A.). S.-S. Lee was supported by NRF grant NRF-2016R1C1B2006697. GYZ is supported by the Korea Research Fellowship Program through the NRF (NRF-2015H1D3A1066561). M.K. acknowledges the financial support of JSPS KAKENHI Grant Numbers JP18K03656 and JP18H03721. J. W. Lee is grateful for the support of the National Research Council of Science and Technology, Korea (Project Number EU-16-001).


\end{document}